\documentclass[journal=nalefd,manuscript=letter,layout=traditional]{achemso}

\usepackage{xcolor} 
\usepackage{textgreek} 
\usepackage{chemformula} 
\usepackage{siunitx} 

\newcommand{\mbapb}{\ch{MBA2PbI4}}
\newcommand{\smbapb}{(\textit{S}-\ch{MBA)2PbI4}}
\newcommand{\rmbapb}{(\textit{R}-\ch{MBA)2PbI4}}
\newcommand{\racmbapb}{(\textit{rac}-\ch{MBA)2PbI4}}
\newcommand{\ba}{\ch{BA2PbI4}}
\newcommand{\pea}{\ch{PEA2PbI4}}

\newcommand{\triplescrew}{$P2_{1}2_{1}2_{1}$}

\newcommand{\vgroup}[1]{\Bar{v}^{\text{Γ}-\text{#1}}}

\newcommand{\phononcircpolgeneral}{s^{\alpha}_{\mathbf{q},\sigma}}

\newcommand{\kB}{k_{\mathrm{B}}}

\author{Mike Pols}
    \affiliation{Materials Simulation \& Modelling, Department of Applied Physics and Science Education, Eindhoven University of Technology, 5600 MB, Eindhoven, The Netherlands}
    \email{m.c.w.m.pols@tue.nl}
\author{Geert Brocks}
    \affiliation{Materials Simulation \& Modelling, Department of Applied Physics and Science Education, Eindhoven University of Technology, 5600 MB, Eindhoven, The Netherlands}
    \alsoaffiliation{Computational Chemical Physics, Faculty of Science and Technology and MESA+ Institute for Nanotechnology, University of Twente, 7500 AE, Enschede, The Netherlands}
\author{Sof\'{i}a Calero}
    \affiliation{Materials Simulation \& Modelling, Department of Applied Physics and Science Education, Eindhoven University of Technology, 5600 MB, Eindhoven, The Netherlands}
\author{Shuxia Tao}
    \affiliation{Materials Simulation \& Modelling, Department of Applied Physics and Science Education, Eindhoven University of Technology, 5600 MB, Eindhoven, The Netherlands}
    \email{s.x.tao@tue.nl}

\title{Chiral Phonons in 2D Halide Perovskites}

\begin{document}

\begin{abstract}
Phonons in chiral crystal structures can be circularly polarized, making them chiral. Chiral phonons carry angular momentum, which is observable in heat currents, and, via coupling to electron spin, in spin currents. Two-dimensional (2D) halide perovskites, versatile direct band gap semiconductors, can easily form chiral structures by incorporating chiral organic cations. As a result, they exhibit phenomena such as chirality-induced spin selectivity (CISS) and the spin Seebeck effect, although the underlying mechanisms remain unclear. Using on-the-fly machine-learning force fields trained against density functional theory calculations, we confirm the presence of chiral phonons, a potential key factor for these effects. Our analysis reveals that low-energy phonons, originating from the inorganic framework, primarily exhibit chirality. Under a temperature gradient, these chiral phonons generate substantial angular momentum, leading to experimentally observable effects. These findings position chiral 2D perovskites as a promising platform for exploring the interplay between phononic, electronic, spintronic, and thermal properties.
\end{abstract}

Chirality, a fundamental property of matter, manifests itself across disciplines, from biology to optical and quantum materials. Chiral molecules or crystals exist in mirror-image forms that cannot be superimposed, giving rise to distinctive properties such as optical rotation and circular dichroism (CD), observed in materials such as tellurium (\ch{Te})~\cite{adesOpticalActivityTellurium1975, ben-mosheEnantioselectiveControlLattice2014}. Beyond optical properties, chirality is also found to affect the properties of charge carriers, i.e. electrons and holes, in materials, for instance in chirality-induced spin selectivity (CISS)~\cite{rayAsymmetricScatteringPolarized1999}, fermions in graphene~\cite{katsnelsonChiralTunnellingKlein2006}, Weyl semimetals~\cite{xuDiscoveryWeylFermion2015}, and topological insulators \cite{hsiehTunableTopologicalInsulator2009}. Moreover, chirality also emerges in bosons, as confirmed by the discovery and characterization of chiral phonons in two-dimensional (2D) transition metal dichalcogenides (TMDs)~\cite{zhangChiralPhononsHighSymmetry2015, zhuObservationChiralPhonons2018}, Moir\'{e} superlattices~\cite{suriChiralPhononsMoire2021, maityChiralValleyPhonons2022} and three-dimensional (3D) materials like α-\ch{HgS}~\cite{ishitoTrulyChiralPhonons2023}, α-\ch{SiO2}~\cite{uedaChiralPhononsQuartz2023}, and \ch{Te}~\cite{zhangWeylPhononsChiral2023}.

Chiral phonons have a nonzero angular momentum~\cite{zhangChiralPhononsHighSymmetry2015}. Under thermal equilibrium the phonon occupancies are such that their total angular momentum averages to zero. However, in heat transport experiments out-of-equilibrium distributions of chiral phonons are created, enabling them to carry an observable amount of angular momentum~\cite{hamadaPhononAngularMomentum2018}, analogous to the Edelstein effect in electrical transport~\cite{edelsteinSpinPolarizationConduction1990, sanchezSpinchargeConversionUsing2013, yodaCurrentinducedOrbitalSpin2015}. This nonzero phonon angular momentum can be experimentally measured as the crystal displays a recoil rotational motion, the Einstein-de Haas effect~\cite{einsteinExperimentellerNachweisAmpereschen1915, einsteinExperimentalProofExistence1915, zhangAngularMomentumPhonons2014}, or via the magnetic moment associated with the angular momentum~\cite{juraschekOrbitalMagneticMoments2019}. Alternatively, the phonon angular momentum can couple with electron spin~\cite{juraschekPhonomagneticAnalogsOptomagnetic2020}, leading to the generation and detection of spin currents~\cite{oheChiralityInducedSelectivityPhonon2024}, as observed in the spin Seebeck effect~\cite{uchidaObservationSpinSeebeck2008, kimChiralphononactivatedSpinSeebeck2023}.

Despite significant advances, chiral effects in crystals are typically dictated by rigid crystal structures, making them difficult to manipulate. Hybrid organic-inorganic 2D halide perovskites form a class of materials with unprecedented tunability, as their properties can be adjusted by substituting different ions. By incorporating chiral organic cations, a chiral crystal structure is formed~\cite{pietropaoloRationalizingDesignImplementation2022, moroniChiral2DQuasi2D2024}, due to a transfer of structural chirality to the metal halide framework~\cite{janaOrganicinorganicStructuralChirality2020, sonUnravelingChiralityTransfer2023, polsTemperatureDependentChiralityHalide2024}. As a result, these materials exhibit remarkable chiroptical properties~\cite{vanormanChiralLightMatter2025}, such as optical rotation and CD~\cite{ahnNewClassChiral2017, apergiCalculatingCircularDichroism2023}, the emission and detection of circularly polarized light~\cite{longSpinControlReduceddimensional2018, yuChiralityInducedSpinOrbit2020, liuBrightCircularlyPolarized2023, yao2023SymmetryBroken2DLead}, and spin-polarized currents without the need for magnetism (CISS)~\cite{luSpindependentChargeTransport2019, kimChiralinducedSpinSelectivity2021}. Furthermore, the suggested link between chiral phonons and the spin Seebeck effect further highlights their potential as a platform for coupling optical, electronic, and thermal properties~\cite{kimChiralphononactivatedSpinSeebeck2023}.

Our work focuses on elucidating the chiral properties of phonons in 2D halide perovskites, a promising yet underexplored domain. Using machine-learning force fields (MLFFs) trained against density functional theory (DFT) calculations~\cite{polsTemperatureDependentChiralityHalide2024}, we characterize the phonon modes of chiral 2D \mbapb{} at the harmonic level. By calculating the angular momentum of phonons as a function of their propagation direction, and mimicking heat transport experiments using the Boltzmann transport equation, we quantify the angular momentum generated under thermal gradients. Our findings demonstrate a significant anisotropy in the generated angular momentum, predominantly within the lead iodide planes, the magnitude of which can be altered by changing the crystal axis along which the gradient is applied. This anisotropic behavior paves the way for tailored spintronic and thermoelectric applications.

\begin{figure*}[htbp!]
    \includegraphics{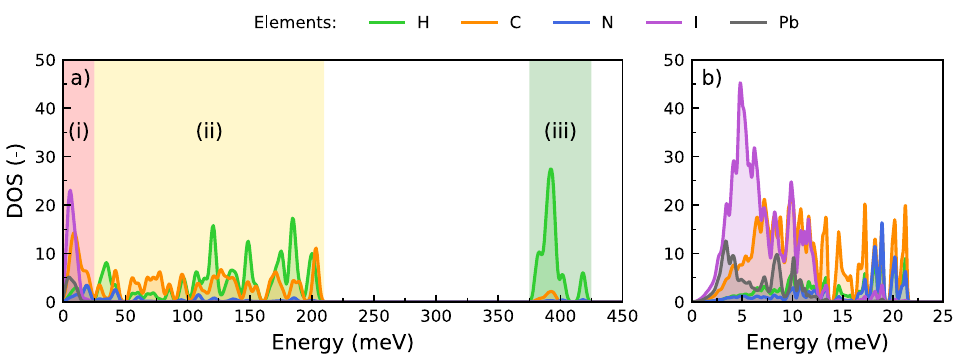}
    \caption{a) Phonon density of states (DOS) of \smbapb{} with the (i) low-energy, (ii) intermediate-energy and (iii) high-energy regions in red, yellow, and green colors, respectively. b) A zoom-in of the low-energy region (0 - \SI{25}{\meV}). Gaussian broadenings of \SI{2.0}{\meV} and \SI{0.1}{\meV} were used in the full and detailed DOS, respectively.}
    \label{fig:phonon_dos} 
\end{figure*}

To assess the character of the lattice vibrations in chiral \smbapb{}, we show its phonon density of states (DOS) in Figure~\ref{fig:phonon_dos}. The details of these calculations can be found in Supporting Notes 1-3. Analogous to three-dimensional (3D) perovskites~\cite{quartiRamanSpectrumCH3NH3PbI32014, brivioLatticeDynamicsVibrational2015}, we can identify three energy regions (Figure~\ref{fig:phonon_dos}a); (i) a low-energy region, $0 - \SI{25}{\meV}$, (ii) an intermediate-energy region, $25-\SI{210}{\meV}$, and (iii) a high-energy region, $375-\SI{425}{\meV}$. Considering the contributions of the different atoms in the crystal lattice, the energy regions can be associated with the motion of different parts of the crystal lattice.

The low-energy region (i), shown in Figure~\ref{fig:phonon_dos}b, is primarily the result of vibrations in the inorganic framework (\ch{[PbI4]^{2-}}), in particular in the lower half of that energy region. Some motion of the organic cations (\ch{MBA+}) is mixed in, especially in the upper half of that energy region, due to coupling between the inorganic framework and the organic molecules~\cite{biegaDynamicDistortionsQuasi2D2023}. The corresponding vibrations involve motions of the cations as a whole. Indeed, comparing the phonon DOS of different 2D perovskites (Supporting Note 4), we find that the contribution of the organic cations at low energies scales with their size and mass.

In contrast, the two higher energy regions are the result of vibrations within the organic cations. The intermediate-energy region (ii) is associated with torsional or bending motion of molecular fragments or, for example, \ch{C}$-$\ch{C} and \ch{C}$-$\ch{N} stretch vibrations. The high-energy modes (iii) correspond to the stretch vibrations of \ch{C}$-$\ch{H} and \ch{N}$-$\ch{H} bonds in the organic cations.

\begin{figure*}[htbp!]
    \includegraphics{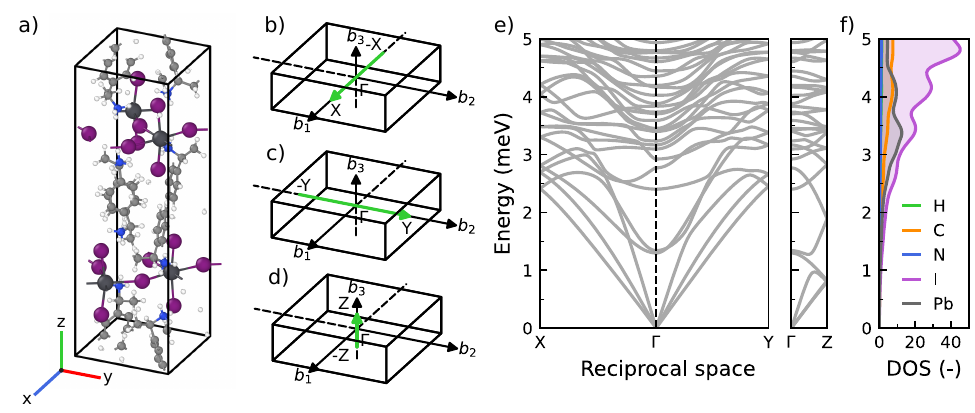}
    \caption{a) Unit cell of \smbapb{} with \triplescrew{} space group. Hydrogen (H), carbon (C), nitrogen (N), iodine (I), and lead (Pb) are represented by white, gray, blue, purple, darkgray spheres, respectively. Brillouin zone of \smbapb{}, with paths along the b) $b_{1}$-axis (Γ$-$X), c) $b_{2}$-axis (Γ$-$Y), and d) $b_{3}$-axis (Γ$-$Z). Special points are X = ($\frac{1}{2}$, 0, 0), Y = (0, $\frac{1}{2}$, 0), and Z = (0, 0, $\frac{1}{2}$), with -X = (-$\frac{1}{2}$, 0, 0). e) Phonon dispersion and f) density of states (DOS) of low-energy region (0 - \SI{5}{\meV}). The DOS is broadened using a Gaussian smearing of \SI{0.1}{\meV}.}
    \label{fig:phonon_dispersion} 
\end{figure*}

In heat transport and electron-phonon coupling the low-energy phonons are particularly relevant. The phonon dispersion in the low-energy region of \smbapb{} (0 - \SI{5}{\meV}) are shown in Figure~\ref{fig:phonon_dispersion}. We specifically focus on the dispersions in the in-plane Γ$-$X and Γ$-$Y as well as out-of-plane Γ$-$Z directions of the inorganic layers within the 2D perovskite (Figure~\ref{fig:phonon_dispersion}a-d). The absence of any imaginary modes in the phonon dispersion indicates that the \triplescrew{} crystal structure of \smbapb{} observed in experiments~\cite{janaOrganicinorganicStructuralChirality2020}, is a stable energy minimum. 

Focusing on the acoustic phonons (Figure~\ref{fig:phonon_dispersion}e-f), we find that in the in-plane directions (Γ$-$X and Γ$-$Y), the three acoustic phonon branches are non-degenerate and  have  different group velocities. In contrast, we observe a near degeneracy between the two TA modes of the acoustic phonons in the out-of-plane direction (Γ$-$Z). Comparing the average group velocities of the acoustic phonons in the three directions ($\vgroup{X}$ = \SI{1861.3}{\m\per\s}, $\vgroup{Y}$ = \SI{1796.3}{\m\per\s}, and $\vgroup{Z}$ = \SI{1669.2}{\m\per\s}), we find that they are actually quite similar, which indicates a surprisingly small anisotropy. We find a similarly low anisotropy in the phonon group velocities of \ba{} and \pea{} in Supporting Note 5, which is in agreement with previous findings~\cite{liRemarkablyWeakAnisotropy2021}.

A phonon eigenmode with wave vector $\mathbf{q}$ and mode index $\sigma$ is described by a polarization vector $\mathbf{e}_{i, \mathbf{q}, \sigma}$, with $i$ labeling the atoms in the unit cell. The polarization vectors are normalized over all atoms of the unit cell, so that $\sum_{i = 1} \mathbf{e}_{i, \mathbf{q}, \sigma}^\dagger \mathbf{e}_{i, \mathbf{q}, \sigma} = 1$. The circular polarization of the phonon modes can be quantified by calculation the phonon circular polarization as
\begin{equation}
    \label{eq:phonon_polarization}
    s^\alpha_{\mathbf{q}, \sigma} =  \sum_{i = 1}^{N} \mathbf{e}_{i, \mathbf{q}, \sigma}^\dagger S^\alpha \mathbf{e}_{i, \mathbf{q}, \sigma},
\end{equation}
where $S^{\alpha}$ $(\alpha = x, y, z$) are the spin-1 matrices on a Cartesian basis. The magnitude and sign of the circular polarization of an eigenmode determine the chirality or handedness of the phonon, with $0 < \phononcircpolgeneral \leq +1$ and $-1 \leq \phononcircpolgeneral < 0$ indicating a right- and left-handed phonon mode, respectively. Achiral phonon modes, with a linear polarization, such as longitudinal modes, have no circular polarization ($\phononcircpolgeneral = 0$).

The circular polarization can, in principle, be measured with respect to any arbitrary axis $\alpha$. However, some components will be zero for symmetry reasons. For instance, in the crystal structure of \smbapb{}, space group \triplescrew{}, for phonons propagating in a direction along one of the crystal axes $\mathbf{q}$ $=$ $(q_x,0,0)$, $(0,q_y,0)$ or $(0,0,q_z)$, only the corresponding $x$-, $y$-, or $z$-component of $s^\alpha_{\mathbf{q}, \sigma}$ is nonzero. In the current work we focus on those phonons propagating either in $x$-, $y$-, or $z$-directions. To calculate $s^\alpha_{\mathbf{q},\sigma}$ we follow the procedures outlined in earlier work~\cite{zhangChiralPhononsHighSymmetry2015,chenChiralPhononDiode2022}. Additional details are provided in Supporting Note 6.

\begin{figure*}[htbp!]
    \includegraphics{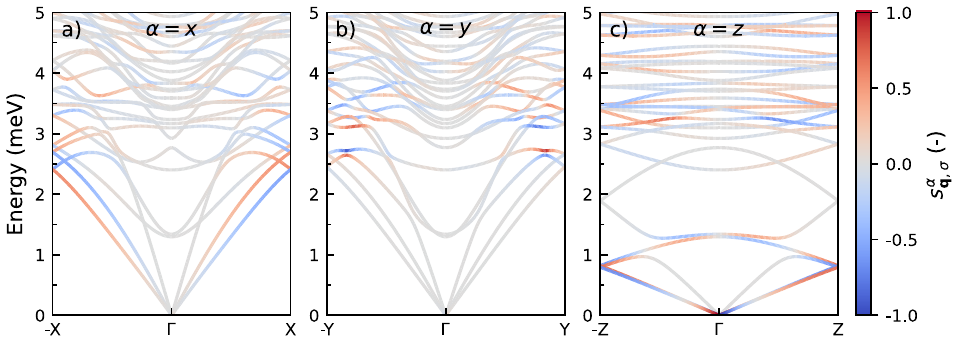}
    \caption{Phonon dispersion along the a) $x$-axis (-X$-$Γ$-$X), b) $y$-axis (-Y$-$Γ$-$Y), and c) $z$-axis (-Z$-$Γ$-$Z) of \smbapb{}. Phonon branches are color-coded with the circular polarization of the phonon modes. Red, blue, and gray are used to represent right-handed ($\phononcircpolgeneral > 0$), left-handed ($\phononcircpolgeneral < 0$), and non-polarized  ($\phononcircpolgeneral = 0$) phonon modes.}
    \label{fig:phonon_polarization_directions} 
\end{figure*}

The dispersion of phonons in the low-energy region of \smbapb{}, propagating along the $x$-, $y$-, and $z$-axis, as well as their respective chirality, is shown in Figure~\ref{fig:phonon_polarization_directions}. Phonons propagating in the positive and negative direction along the different axes are shown in separate panels. Figure~\ref{fig:phonon_polarization_directions} demonstrates that in all directions chiral phonons can be found. One observes that phonons propagating in opposite directions, for instance, Γ$-$X and -X$-$Γ, see (Figure~\ref{fig:phonon_polarization_directions}a), have an opposite circular polarization, i.e. $s^x_{\mathbf{q}, \sigma} = - s^x_{-\mathbf{q}, \sigma}$. This is a consequence of time-reversal symmetry~\cite{hamadaPhononAngularMomentum2018}, and is similar to the relation between spin-orbit split bands in an electronic band structure. Indeed, a similar coupling between the phonon propagation direction and its polarization has been observed in both \ch{Te} and α-\ch{SiO2}~\cite{chenChiralPhononDiode2022}.

By examining the chirality of the phonons across the whole spectrum, we find that the chiral phonon modes are primarily found in the low-energy region of the phonon spectrum of \smbapb{}. In this region (0 - \SI{25}{\meV}), phonons can possess a substantial chirality ($|\phononcircpolgeneral| \geq \frac{1}{5}$), whereas that of higher energy phonons ($>$ \SI{25}{\meV}) is negligible, as shown in Figure S8 in Supporting Note 6. Generally speaking, the phonon chirality appears to increase for phonons with wave vectors approaching the zone boundaries. In the $x$-direction (Figure~\ref{fig:phonon_polarization_directions}a), the two lowest acoustic branches show appreciable chirality, as do several of the low-energy optical branches. Chirality in the $y$-direction is predominantly observed in the optical modes near the zone boundary (Figure~\ref{fig:phonon_polarization_directions}b). The two lowest acoustic modes in the $z$-direction seem to show appreciable chirality (Figure~\ref{fig:phonon_polarization_directions}c), but this is slightly misleading, as they are almost degenerate, and their chirality sums up to zero. 

As mentioned earlier, the low-energy phonons are those phonons which heavily involve motions of the atoms within the inorganic framework. In previous work, we established that this inorganic framework has chiral structural distortions, resulting from the transfer of chirality from the organic cations to the inorganic framework~\cite{polsTemperatureDependentChiralityHalide2024}. In contrast, achiral 2D perovskites were found to lack such chiral distortions and chiral phonon modes. Indeed, for achiral 2D perovskites, i.e. \racmbapb{}, \ba{}, or \pea{}, we do not observe any circular polarization of the phonon modes, as shown in Supporting Note 6. 

To support the evidence of the relation between structural chirality and phonon chirality, we have also compared the phonons in the two enantiomers of \mbapb{} in Supporting Note 7. Whereas the phonon dispersions are identical for the two enantiomers, we observe that phonons propagating in the same direction in each enantiomer have opposite polarization; for each phonon branch, a right-handed phonon in \smbapb{} becomes a left-handed phonon in \rmbapb{}, and vice versa. We propose that the chiral distortions within the inorganic framework~\cite{polsTemperatureDependentChiralityHalide2024}, which can readily be tuned through compositional engineering~\cite{luHighlyDistortedChiral2020, sonUnravelingChiralityTransfer2023}, play a crucial role in the emergence of circularly polarized phonons in 2D perovskites.

\begin{figure*}[htbp!]
    \includegraphics{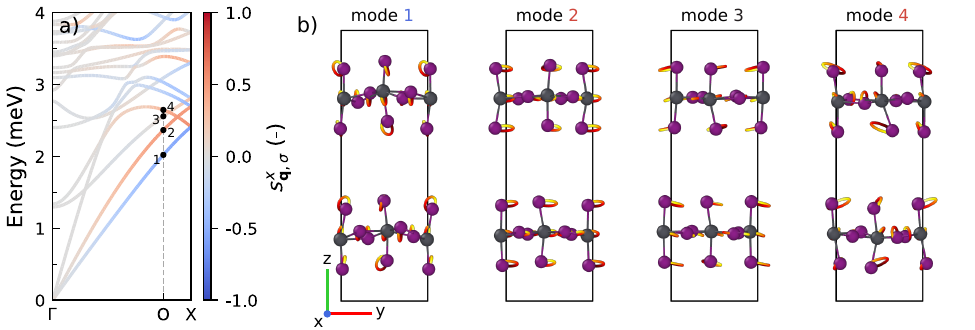}
    \caption{a) Phonon dispersion and b) atomic motion in the selected phonon modes at O = ($\frac{2}{5}$, 0, 0) in \smbapb{}. All mode numbers are colored to indicate the circular polarization, with red (right-handed), black (non-polarized), and blue (left-handed). The atoms follow the trajectories from red to yellow as time progresses.}
    \label{fig:phonon_atomic_motion} 
\end{figure*}

Figure~\ref{fig:phonon_atomic_motion} shows an example of the atomic motions of chiral phonons in \smbapb{}. It illustrates the atomic motion of the lowest four phonon modes at the point O = ($\frac{2}{5}$, 0, 0) along the in-plane Γ$-$X path. Modes 1, 2, and 4 are chiral, whereas mode 3 is achiral. The chiral modes exhibit an elliptical motion in the $yz$-plane, perpendicular to their propagation direction ($x$-direction). The achiral mode exhibits a linear oscillatory motion within this plane. Comparing the two lowest phonon modes, 1 and 2, we see that the semi-major axis of the elliptical motion for these two modes has a different orientation; mode 1 has it oriented along the $z$-axis, whereas it is parallel to the $y$-axis for mode 2. A similar analysis can be found in Supporting Note 7 for the phonons propagating along the Γ$-$Y and Γ$-$Z paths.

Having predicted the presence of chiral phonons in 2D halide perovskites, one needs a way to establish them experimentally. Figure~\ref{fig:phonon_polarization_directions} demonstrates that in every phonon branch $\sigma$ phonons moving in opposite directions, i.e. $\pm \mathbf{q}$, exhibit opposite chirality, such that $s^{\alpha}_{\mathbf{q}, \sigma} = - s^{\alpha}_{-\mathbf{q}, \sigma}$. In thermal equilibrium, their Bose-Einstein occupation numbers $f_{0}\left( \omega_{\mathbf{q}, \sigma} \right)$ are equal, since for every mode $\omega_{\mathbf{q}, \sigma}=\omega_{\mathbf{-q}, \sigma}$, which means their total chirality sums to zero.

However, the phonon distribution can be driven out-of-equilibrium by applying a temperature gradient along arbitrary directions, which generates a heat flux. This breaks the symmetry between occupations of the right- and left-moving modes, i.e. $f\left( \omega_{\mathbf{q}, \sigma} \right) \neq f\left( \omega_{-\mathbf{q}, \sigma} \right)$, and generates a phonon distribution with a nonzero net chirality. The effect is analogous to the Edelstein effect in electronic transport. As established earlier, phonon chirality is defined by a nonzero phonon circular polarization (Eqn.~\ref{eq:phonon_polarization}). As the latter is a form of angular momentum, it is possible to measure it, either directly, using the Einstein-de Haas effect~\cite{zhangObservationPhononAngular2024}, or indirectly, via a coupling between phonons and magnetic moments, and the inverse spin-Hall effect~\cite{oheChiralityInducedSelectivityPhonon2024}.

To calculate the angular momentum, we follow the procedure as formulated by \citet{hamadaPhononAngularMomentum2018}. Heat transport is described with the Boltzmann transport equation, which is linearized under the assumption of a sufficiently small temperature gradient. In principle, each phonon mode ($\mathbf{q},\sigma$) has its own lifetime $\tau_{\mathbf{q},\sigma}$. However, calculating these individual lifetimes, for instance by accounting for anharmonic scattering processes, is currently computationally infeasible for a system of this size. Then it is common practice to approximate $\tau_{\mathbf{q},\sigma}$ with a single uniform relaxation time $\tau$~\cite{hamadaPhononAngularMomentum2018}. Under these conditions, the components of the angular momentum are given by
\begin{equation}
    \label{eq:volumetric_angular_momentum_generation}
    J^{\mathrm{ph},\alpha} = - \frac{\hbar\tau}{V} \sum_{\mathbf{q}, \sigma; \beta=x,y,z} s^{\alpha}_{\mathbf{q}, \sigma} v^\beta_{\mathbf{q}, \sigma} \frac{\partial f_{0} \left( \omega_{\mathbf{q}, \sigma} \right)}{\partial T} \frac{\partial T}{\partial x^{\beta}} \equiv \sum_{\beta} \alpha^{\alpha \beta} \frac{\partial T}{\partial x^{\beta}},
\end{equation}
where $J^{\mathrm{ph},\alpha}$ ($\alpha = x, y, z$) are the components of the total phonon angular momentum per unit volume, $\hbar$ is the reduced Planck constant, $\tau$ is the phonon relaxation time, $V$ the unit cell volume, $s^{\alpha}_{\mathbf{q}, \sigma}$ the phonon circular polarization, $v^\beta_{\mathbf{q}, \sigma}$ ($\beta=x,y,z$) the components of the phonon group velocity, and $f_{0}$ the Bose-Einstein distribution. The response tensor of the material is then defined by $\alpha^{\alpha \beta}$. It is a second rank tensor obeying the symmetry rules of the crystal structure, which in case of space group \triplescrew{} (point group $D_{2}$) gives off-diagonal elements of zero, $\alpha^{\alpha \beta}$ $=$ 0 ($\alpha \neq \beta$), and unequal diagonal elements, $\alpha^{xx} \neq \alpha^{yy} \neq \alpha^{zz}$ \cite{hamadaPhononAngularMomentum2018}. A more detailed discussion on the calculation of this quantity can be found in Supporting Note 8, as well as details on the convergence of the results.

\begin{figure*}[htbp]
    \includegraphics{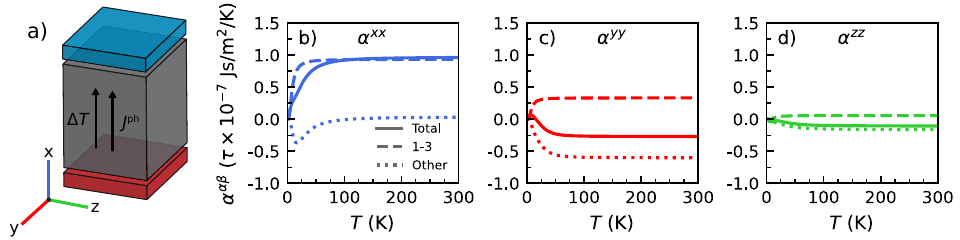}
    \caption{a) Illustration of the induced angular momentum from phonons ($J^{\mathrm{ph}}$) in \smbapb{} for a temperature gradient ($\Delta T$) in the $x$-direction. The temperature dependence of the b) $\alpha^{xx}$, c) $\alpha^{yy}$, and d) $\alpha^{zz}$ components of the response tensor given by solid lines. The contributions from the three lowest energy phonons and the other phonons are given by dashed and dotted lines, respectively.}
    \label{fig:angular_momentum_generation} 
\end{figure*}

Based upon the calculated spectrum of phonon modes in \smbapb{} and their chirality, we have calculated the induced angular momentum density according to Eqn.~\ref{eq:volumetric_angular_momentum_generation} as a function of temperature. The results are shown in Figure~\ref{fig:angular_momentum_generation}. Because of the symmetry of the structure, the induced angular momentum is parallel to the applied temperature gradient. In all directions, we observe that a gradient in the temperature results in the generation of a nonzero angular momentum. At low temperatures ($<$ \SI{150}{\K}),  the angular momentum shows a strong dependency on temperature, but for higher temperatures ($>$ \SI{150}{\K}), it becomes independent of the temperature. This merely confirms that the low-energy phonons in \smbapb{} cause the chiral effect, as for $\kB T \gg \hbar \omega_{\mathbf{q}, \sigma}$, $f_{0} \left( \omega_{\mathbf{q}, \sigma} \right) \propto T$, so $J^{\mathrm{ph},\alpha}$ should become independent of the temperature.

At \SI{300}{\K}, the induced angular momentum is largest for a temperature gradient applied in the $x$-direction ($\alpha^{xx} = +9.6 \cdot \tau \times10^{-8}$ \SI{}{\J\s\per\m\squared\per\K}), with gradients applied in the $y$- and $z$-directions showing markedly lower responses and with an opposite sign ($\alpha^{yy} = -2.7 \cdot \tau \times10^{-8}$ \SI{}{\J\s\per\m\squared\per\K} and $\alpha^{zz} = -1.1 \cdot \tau \times10^{-8}$ \SI{}{\J\s\per\m\squared\per\K}). In the current model, the angular momenta (Eq. \ref{eq:volumetric_angular_momentum_generation}) and the elements of the response tensor ($\alpha^{\alpha \beta}$) are linearly dependent on the relaxation time $\tau$. A conservative estimate of the latter ($\tau \approx \SI{1}{\ps}$)~\cite{acharyyaIntrinsicallyUltralowThermal2020} yields an angular momentum that is large enough to be experimentally observable~\cite{hamadaPhononAngularMomentum2018}, particularly along the $x$-direction, either through conversion into spin signals~\cite{oheChiralityInducedSelectivityPhonon2024} or via torque measurements~\cite{zhangObservationPhononAngular2024}.

Near the Brillouin zone (BZ) center at Γ the three lowest energy branches are formed by the acoustic modes, but near the BZ edges these hybridize with the lowest energy optical modes, so the distinction between acoustic and optical modes is blurred there, see Figure~\ref{fig:phonon_polarization_directions}. Nevertheless, it is instructive to decompose the response tensors into contributions from the three lowest energy phonons and the rest. We find that above \SI{100}{\K} the contributions to $\alpha^{xx}$ almost completely come from the three lowest energy phonons (Figure~\ref{fig:angular_momentum_generation}b), with the higher energy optical phonons having essentially no effect. In contrast, for $\alpha^{yy}$ and $\alpha^{zz}$ (Figure~\ref{fig:angular_momentum_generation}c-d), the three lowest energy phonons and the optical phonons have similar contributions. These have opposite signs however, which explains the smaller values of $\alpha^{yy}$ and $\alpha^{zz}$ as compared to $\alpha^{xx}$, since they cancel out.

According to Eqn.~\ref{eq:volumetric_angular_momentum_generation}, a phonon mode should have both a high chirality and high velocity to generate a substantial angular momentum. Therefore, it is not so surprising that the chiral acoustic modes provide a dominant contribution. As can be seen in Figure~\ref{fig:phonon_polarization_directions}, these can be found along the Γ$-$X path, but not along the Γ$-$Y path, whereas along the Γ$-$Z path, modes of opposite chirality are almost degenerate, canceling their contributions.

In summary, we investigated the vibrational properties of chiral 2D perovskites, using \mbapb{} as a representative example. Our findings highlight that the low-energy vibrational modes (0 - \SI{25}{\meV}) are primarily associated with the inorganic framework. In contrast, the intermediate-energy (25 - \SI{210}{\meV}) and high-energy (375 - \SI{425}{\meV}) vibrations, are associated with movement of the organic cations.

A key result of our analysis is the identification of chiral phonons in the low-energy vibrational spectrum. The handedness of the chiral phonons is directly coupled to both their propagation direction and the structural chirality of the crystal. Reversing either parameter results in a corresponding reversal of the phonon handedness. This coupling enables chiral 2D perovskites to generate observable angular momentum under a temperature gradient, unveiling new functionality for these materials. Notably, our results reveal pronounced anisotropy, with angular momentum primarily generated within the lead iodide planes. This angular momentum can be modulated by adjusting the crystal axis along which the temperature gradient is applied, suggesting new possibilities for directionally controlled spintronic and thermoelectric applications.

Our work provides exciting opportunities to explore the intricate interplay between phononic, electronic, spintronic, and optical phenomena. The compositional tunability of these materials, achieved through the substitution of metal ions, halide ions, or organic cations, offers unprecedented control over their structure and the behavior of the phonons. By linking their structural chirality to vibrational, electronic, and thermal properties, these materials enable the study of complex phenomena such as chirality-induced spin selectivity (CISS), the spin Seebeck effect, and other emergent behavior. These insights deepen our understanding of fundamental chiral processes and pave the way for applications in spintronics, thermoelectrics, and quantum materials, where precise control over angular momentum and phonon polarization is crucial.

\begin{suppinfo}

\begin{itemize}
    \item Computational details of density functional theory (DFT) calculations, machine-learning force field (MLFF) training, and phonon calculations; comparison of phonon density of states (DOS); phonon group velocities; circular polarization of phonons; atomic motion of phonon modes; phonon angular momentum response tensor 
\end{itemize}

\end{suppinfo}

\begin{acknowledgement}
S.T. acknowledges funding from Vidi (project no. VI.Vidi.213.091) from the Dutch Research Council (NWO).
\end{acknowledgement}

\clearpage

\bibliography{ms}

\end{document}


\title{Supporting Information: \\ Chiral Phonons in 2D Halide Perovskites}

\author{Mike Pols}
\email{m.c.w.m.pols@tue.nl}
\affiliation{%
    Materials Simulation \& Modelling, Department of Applied Physics and Science Education, Eindhoven University of Technology, 5600 MB, Eindhoven, The Netherlands
}%

\author{Geert Brocks}
\affiliation{%
    Materials Simulation \& Modelling, Department of Applied Physics  and Science Education, Eindhoven University of Technology, 5600 MB, Eindhoven, The Netherlands
}%
\affiliation{%
    Computational Chemical Physics, Faculty of Science and Technology and MESA+ Institute for Nanotechnology, University of Twente, 7500 AE, Enschede, The Netherlands
}%

\author{Sof\'{i}a Calero}
\affiliation{%
    Materials Simulation \& Modelling, Department of Applied Physics and Science Education, Eindhoven University of Technology, 5600 MB, Eindhoven, The Netherlands
}%

\author{Shuxia Tao}
\email{s.x.tao@tue.nl}
\affiliation{%
    Materials Simulation \& Modelling, Department of Applied Physics and Science Education, Eindhoven University of Technology, 5600 MB, Eindhoven, The Netherlands
}%



\maketitle

\clearpage

\tableofcontents

\clearpage

\section{SI Note: Density functional theory (DFT)}

Density functional theory (DFT) calculations were performed using the Vienna Ab-initio Simulation Package (\texttt{VASP})~\cite{kresseInitioMoleculardynamicsSimulation1994, kresseEfficiencyAbinitioTotal1996, kresseEfficientIterativeSchemes1996}. The projector-augmented wave (PAW) pseudopotentials~\cite{kresseUltrasoftPseudopotentialsProjector1999} had the following valence electron configurations: \ch{H} (1s\textsuperscript{1}), \ch{C} (2s\textsuperscript{2}2p\textsuperscript{2}), \ch{N} (2s\textsuperscript{2}2p\textsuperscript{3}), \ch{I} (5s\textsuperscript{2}5p\textsuperscript{5}) and \ch{Pb} (6s\textsuperscript{2}6p\textsuperscript{2}). In the machine-learning force fields (MLFF) the electronic exchange-correlation interactions were modeled with the meta-GGA SCAN functional~\cite{sunStronglyConstrainedAppropriately2015}. For comparison, we also trained an MLFF against the Perdew, Burke and Ernzerhof (PBE) functional~\cite{perdewGeneralizedGradientApproximation1996} with long range dispersion interactions accounted for by the DFT-D3(BJ) scheme~\cite{grimmeEffectDampingFunction2011}. In all calculations, an energy cutoff of \SI{500}{\eV} was used. A Γ-centered \kpoints{2}{2}{1} $\mathbf{k}$-mesh, with the single $\mathbf{k}$-point in the direction perpendicular to the inorganic layers, was used for DFT calculations of \smbapb{}, \ba{} and \pea{}. As such, a Γ-centered \kpoints{1}{2}{2} $\mathbf{k}$-mesh was used for \racmbapb{}. 

As a starting point for the calculations, we used experimental crystal structures. An overview of the structures used is found in Table~\ref{tab:perovskite_structures}. All structures were optimized by allowing the atomic positions, cell shape, and cell volume to change, using an energy and force convergence criterion of 10\textsuperscript{-5} \SI{}{\eV} and 10\textsuperscript{-2} \SI{}{\eV\per\angstrom}, respectively. The optimized crystal geometries and the used $\mathbf{k}$-meshes are shown in Table~\ref{tab:perovskites_optimized}, demonstrating good agreement between the DFT calculations and the experimental geometries.

\begin{table}[htbp]
    \caption{Experimental crystal structures of investigated 2D halide perovskites.}
    \label{tab:perovskite_structures}
    \begin{ruledtabular}
    \begin{tabular}{cccccc}
        Perovskite                  & Space group                       & $N_{\mathrm{group}}$  & $T_{\mathrm{exp.}}$ (\SI{}{\K})   & Database ID                          & References                                                                    \\ \hline
        \smbapb{}                   & \triplescrew                      & 19                    & 298                               & \ccdc{2015617}                       & Ref.\citep{janaOrganictoinorganicStructuralChirality2020}                     \\
        \racmbapb{}                 & \centroscrew{}                    & 14                    & 293                               & \ccdc{1877052}                       & Ref.\citep{dangBulkChiralHalide2020}                                          \\ \hline
        \ba{}                       & \pbca                             & 61                    & 100                               & \ccdc{2018893}                       & Ref.\citep{menahemStronglyAnharmonicOctahedral2021}                           \\
        \pea{}                      & \ponebar{}                        & 2                     & 296                               & \ccdc{1542461}                       & Ref.\citep{duTwoDimensionalLeadII2017}                                        \\
    \end{tabular}
    \end{ruledtabular}
\end{table}

\begin{sidewaystable}[htbp]
    \caption{Experimental, DFT-optimized, and MLFF-optimized crystal geometries for 2D halide perovskites.}
    \label{tab:perovskites_optimized} 
    \begin{ruledtabular}
    \begin{tabular}{cccccccccc}
        Perovskite                  & Type          & $a$ (\SI{}{\angstrom}) & $b$ (\SI{}{\angstrom}) & $c$ (\SI{}{\angstrom}) & $\alpha$ (\SI{}{\degree}) & $\beta$ (\SI{}{\degree}) & $\gamma$ (\SI{}{\degree}) & $V$ (\SI{}{\angstrom\cubed}) & $\mathbf{k}$-mesh \\ \hline
        \multirow{3}{*}{\smbapb}    & Exp.          &  8.90 &  9.31 & 28.86 &  90.0 &  90.0 &  90.0 & 2393.3 & \\
                                    & SCAN          &  8.87 &  9.19 & 28.76 &  90.0 &  90.0 &  90.0 & 2342.5 & \kpoints{2}{2}{1} \\
                                    & MLFF          &  8.87 &  9.16 & 28.77 &  90.0 &  90.0 &  90.0 & 2337.6 & \\ \hline
        \multirow{3}{*}{\racmbapb}  & Exp.          & 14.62 &  9.38 &  8.78 &  90.0 & 100.1 &  90.0 & 1185.7 & \\
                                    & SCAN          & 14.66 &  9.35 &  8.66 &  90.0 & 100.8 &  90.0 & 1167.0 & \kpoints{1}{2}{2} \\
                                    & MLFF          & 14.57 &  9.29 &  8.61 &  90.0 &  99.0 &  90.0 & 1151.5 & \\ \hline
        \multirow{3}{*}{\ba}        & Exp.          &  8.42 &  9.00 & 26.08 &  90.0 &  90.0 &  90.0 & 1975.6 & \\
                                    & SCAN          &  8.45 &  9.00 & 26.10 &  90.0 &  90.0 &  90.0 & 1986.6 & \kpoints{2}{2}{1} \\
                                    & MLFF          &  8.41 &  9.01 & 26.33 &  90.0 &  90.0 &  90.0 & 1994.9 & \\ \hline
        \multirow{3}{*}{\pea}       & Exp.          &  8.74 &  8.74 & 33.00 &  84.6 &  84.7 &  89.6 & 2498.3 & \\
                                    & SCAN          &  8.70 &  8.72 & 32.76 &  85.6 &  85.7 &  89.4 & 2471.5 & \kpoints{2}{2}{1} \\
                                    & MLFF          &  8.71 &  8.74 & 32.59 &  85.4 &  85.3 &  89.4 & 2467.3 & \\
    \end{tabular}
    \end{ruledtabular}
\end{sidewaystable}

\clearpage

\section{SI Note: Machine-learning force fields (MLFFs)}

Machine-learning force fields (MLFFs) were trained against total energies, forces, and stresses from density functional theory (DFT) calculations. The training sets were automatically constructed using an on-the-fly active learning scheme together with dynamical simulations in an \textit{NpT} ensemble~\cite{jinnouchiOntheflyMachineLearning2019, jinnouchiPhaseTransitionsHybrid2019}. Local atomic environments were described using an adaptation of the smooth overlap of atomic positions (SOAP) descriptor~\cite{bartokRepresentingChemicalEnvironments2013}, for which we employed a cutoff for the two-body radial descriptor of \SI{6.0}{\angstrom} and a cutoff of \SI{4.0}{\angstrom} for the three-body angular descriptor. The atomic positions were broadened using Gaussian distributions with a width of \SI{0.5}{\angstrom}. Both descriptors were expanded on a basis set of spherical Bessel functions and Legendre polynomials, using 8 and 6 Bessel functions for the two-body and three-body descriptors, respectively, with a maximum quantum number of angular momentum of $l_{\mathrm{max}}$ = 2. The expansion coefficients of this basis set constituted the descriptor for the local atomic environments. To measure the similarity between two local atomic environments, a polynomial kernel function to power 4 was used, in which the two-body radial and three-body angular descriptor vectors were weighted by 0.1 and 0.9, respectively.

The MLFF training was initialized with a constant temperature simulation at \SI{300}{\K}, starting from the optimized crystal structure. This training run was followed by consecutive constant temperature simulations at \SI{100}{\K} and \SI{450}{\K}, each using the final positions and velocities of the previous run as the starting point. All constant temperature simulations were \SI{50}{\ps}. In the final training run, the system was cooled from \SI{350}{\K} to \SI{50}{\K} over \SI{60}{\ps}. The starting point for this simulation, i.e. atomic positions and velocities, was obtained using \SI{10}{\ps} equilibration runs with the intermediate MLFFs. During training the temperature and pressure were controlled using Parrinello-Rahman dynamics~\cite{parrinelloCrystalStructurePair1980, parrinelloPolymorphicTransitionsSingle1981} using friction coefficients $\gamma$ = \SI{5}{\per\ps} and $\gamma_{\mathrm{L}}$ = \SI{5}{\per\ps} for the atomic and lattice degrees of freedom, respectively. A time step of $\Delta t$ = \SI{2}{\fs} and a hydrogen mass of $m_{\ch{H}}$ = \SI{4}{\amu} were used to enhance the sampling rate of new structures. To control the size of the force fields, the number of local reference configurations was capped at 2000, allowing configurations to be discarded once this threshold was reached. Using the training data of the final MLFFs, the models were refit onto faster descriptors without any Bayesian error estimation to speed up the evaluation for large-scale molecular dynamics production runs. Four models were trained in total. The size of the training sets for the MLFFs are shown in Supplementary Note 2.
 
In total we trained four machine-learning force fields (MLFFs), of which the training sets and number of local reference configurations are summarized in Table~\ref{tab:mlff_details}. The models show good agreement with DFT calculations in describing the geometries and dynamics of the halide perovskites. Geometries optimized with the MLFFs based on the SCAN XC functional are shown in Table~\ref{tab:perovskites_optimized}. A complete validation of the dynamics of the MLFFs is found in previous work~\cite{polsTemperatureDependentChiralityHalide2024}. We highlight the use of a single model for \mbacation{}-based perovskites, i.e. \smbapb{}, \rmbapb{}, and \racmbapb{}, as we found a high transferability across the various enantiomers.

\begin{table}[htbp]
    \caption{Size of the training set and number of local reference configurations for various machine-learning force fields (MLFFs) for 2D perovskites.}
    \label{tab:mlff_details} 
    \begin{ruledtabular}
    \begin{tabular}{cccccccc}
         \multirow{2}{*}{XC functional} & \multirow{2}{*}{Training structure}   & \multirow{2}{*}{$N_{\mathrm{DFT}}$ (-)}  & \multicolumn{5}{c}{$N_{\mathrm{basis}}$ (-)}                  \\
                                        &                                       &                                          & \ch{H}    & \ch{C}    & \ch{N}    & \ch{I}    & \ch{Pb}       \\ \hline
         \multirow{3}{*}{SCAN}          & \smbapb{}                             & 972                                      & 2000      & 2000      & 563       & 1240      & 310           \\
                                        & \pea{}                                & 1030                                     & 2000      & 2000      & 507       & 1175      & 218           \\
                                        & \ba{}                                 & 973                                      & 2000      & 1983      & 471       & 1129      & 226           \\ \hline
         PBE+D3(BJ)                     & \smbapb{}                             & 933                                      & 2000      & 2000      & 568       & 1190      & 283           \\
    \end{tabular}
    \end{ruledtabular}
\end{table}

\clearpage

\section{SI Note: Phonon calculations}

Harmonic phonon calculations were performed using the finite displacement method in supercells, as implemented in \texttt{phonopy}~\cite{togoFirstprinciplesPhononCalculations2023, togoImplementationStrategiesPhonopy2023}. The crystal structures were optimized with machine-learning force fields (MLFFs), allowing the atomic positions, cell shape, and cell volume to change until the forces are below 10\textsuperscript{-3} \SI{}{\eV\per\angstrom}. For all perovskites, except \racmbapb{}, the interatomic force constants were computed in a \supercell{3}{3}{1} supercell and the phonon density of states (DOS) on a \kpoints{21}{21}{7} Γ-centered $\mathbf{q}$-mesh. The response tensor $\alpha^{\alpha \beta}$ for \smbapb{} was calculated on the same \kpoints{21}{21}{7} Γ-centered $\mathbf{q}$-mesh, following the procedure outlined by \citet{hamadaPhononAngularMomentum2018}. For \racmbapb{}, a \supercell{2}{3}{3} supercell and a \kpoints{14}{21}{21} Γ-centered $\mathbf{q}$-mesh were used for the interatomic force constants and the phonon DOS, respectively.

\begin{figure*}[htbp!]
    \includegraphics{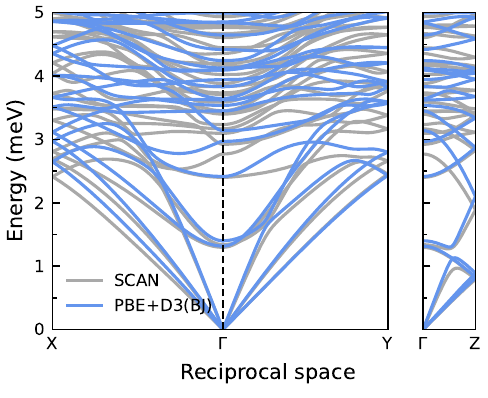}
    \caption{Phonon dispersion of \smbapb{} computed using an MLFF trained against the SCAN and PBE+D3(BJ) XC functionals are shown in gray and blue, respectively.}
    \label{fig:phonon_band_structure_xc} 
\end{figure*}

\begin{figure*}[htbp!]
    \includegraphics{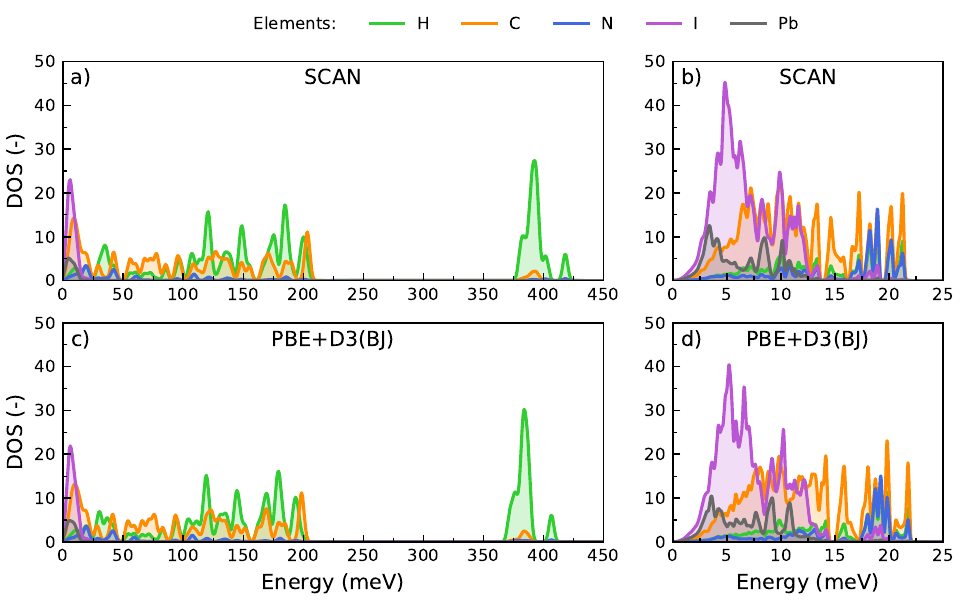}
    \caption{Phonon density of states (DOS) of \smbapb{} using (a-b) SCAN and (c-d) PBE+D3(BJ) exchange-correlation (XC) functional. Gaussian broadenings of \SI{2.0}{\meV} and \SI{0.1}{\meV} were used in the full and detailed DOS, respectively.}
    \label{fig:phonon_dos_xc} 
\end{figure*}

To assess the effect of the type of exchange-correlation (XC) functional on the phonon calculations, we compared the phonon dispersion obtained using MLFFs trained against different XC functionals. In Figure~\ref{fig:phonon_band_structure_xc}, the phonon dispersion of \smbapb{} calculated using an MLFF trained against the SCAN and PBE+D3(BJ) XC functionals are shown. Both MLFFs give qualitatively the same phonon dispersion, with the model trained on PBE+D3(BJ) showing slightly higher energies for equivalent phonon modes. We attribute this shift towards higher energies to differences between the two XC functionals. Identically, we only observe small changes in the phonon density of states between the different functionals, as shown in Figure~\ref{fig:phonon_dos_xc}.

\clearpage

\section{SI Note: Density of states (DOS)}

\subsection{Enantiomeric effects}

\begin{figure*}[htbp!]
    \includegraphics{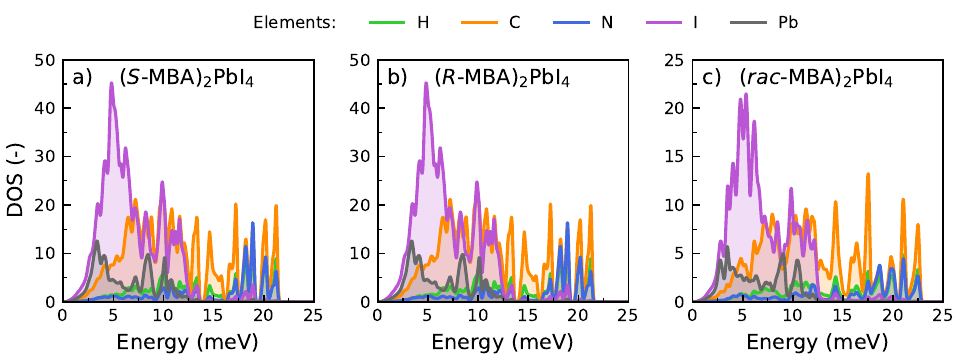}
    \caption{Low energy (0 - \SI{25}{\meV}) phonon density of states (DOS) of the enantiomers of \mbacation{}. Phonon DOS of (a) \smbapb{}, (b) \rmbapb{}, and (c) \racmbapb{}. All DOS were broadened using \SI{0.1}{\meV} of Gaussian broadening.}
    \label{fig:phonon_dos_enantiomers} 
\end{figure*}

To investigate the effect of structural enantiomers of \mbapb{} on the phonon spectrum, particularly the low enery vibrations, we compared the DOS of the enantiomers. In Figure~\ref{fig:phonon_dos_enantiomers} the DOS of \smbapb{}, \rmbapb{}, and \racmbapb{} are shown. The comparison shows that the DOS of the two mirror images, \smbapb{} and \rmbapb{} (Figure~\ref{fig:phonon_dos_enantiomers}a-b), is identical, with \racmbapb{} (Figure~\ref{fig:phonon_dos_enantiomers}c) only showing differences in the fine structure of the DOS. Altogether, this highlights that the structural enantiomers of a 2D halide perovskite exhibit rather similar vibrational characteristics.

\subsection{Influence of cations}

\begin{figure*}[htbp!]
    \includegraphics{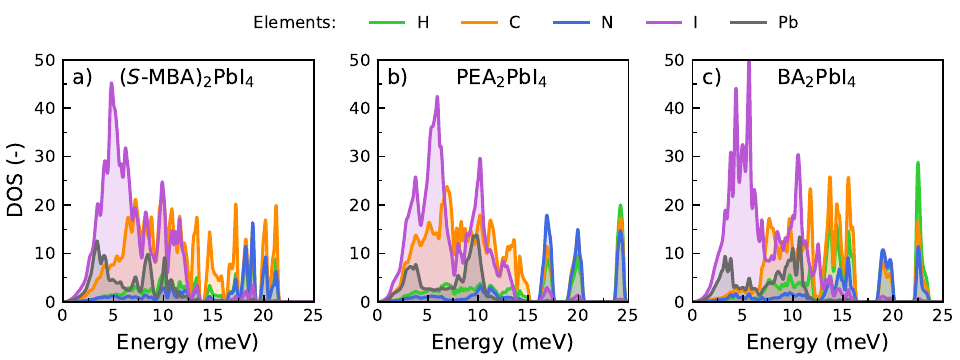}
    \caption{Low energy (0 - \SI{25}{\meV}) phonon density of states (DOS) of various 2D perovskites. Phonon DOS of (a) \smbapb{}, (b) \pea{}, and (c) \ba{}. All DOS were broadened using \SI{0.1}{\meV} of Gaussian broadening.}
    \label{fig:phonon_dos_cations} 
\end{figure*}

To assess the influence of the size and mass of the organic cations on the coupling with the low energy vibrations, we compare the DOS of various 2D perovskites. In Figure~\ref{fig:phonon_dos_cations} the DOS of \smbapb{}, \pea{}, and \ba{} are shown. The comparison shows that large and heavy cations (Figure~\ref{fig:phonon_dos_cations}a-b; $m_{\mbacation{}}$ = \SI{122.191}{\amu} and $m_{\peacation{}}$ = \SI{122.191}{\amu}) couple substantially more with the low energy inorganic framework vibrations than the smaller and lighter cation (Figure~\ref{fig:phonon_dos_cations}c; $m_{\bacation{}}$ = \SI{74.147}{\amu}).

\clearpage

\section{SI Note: Phonon group velocities}

The group velocity of phonons is defined by the following relation
\begin{equation}
    \label{eq:group_velocity}
    \mathbf{v}_{\mathbf{q}, \sigma} = \nabla_{\mathbf{q}} \omega_{\mathbf{q}, \sigma},
\end{equation}
which is the gradient of the angular frequency $\omega_{\mathbf{q}, \sigma}$ as a function of the wave vector $\mathbf{q}$. The group velocity of the acoustic phonons is determined arbitrarily close to the Γ-point along high-symmetry paths. For the 2D halide perovskites, we determine the group velocity for the Γ$-$X, Γ$-$Y, and Γ$-$Z paths. The resulting group velocities of the three acoustic phonons as well as their average value ($\Bar{v}$) are shown in Table~\ref{tab:acoustic_group_velocities}. The investigated 2D perovskites, \smbapb{}, \ba{}, and \pea{}, all exhibit a low anisotropy in the group velocity of the phonons. 

\begin{table}[htbp]
    \caption{Group velocities of the acoustic phonons in various 2D perovskites.}
    \label{tab:acoustic_group_velocities}
    \begin{ruledtabular}
    \begin{tabular}{cccccc}
        \multirow{2}{*}{Perovskite}             & \multirow{2}{*}{Mode} & \multicolumn{3}{c}{Group velocities}                                                                      \\
                                                &                       & $\vg{X}$  (\SI{}{\m\per\s})       & $\vg{Y}$  (\SI{}{\m\per\s})       & $\vg{Z}$  (\SI{}{\m\per\s})       \\ \hline
        \multirow{4}{*}{\smbapb{}}              & 1                     & 1151.1                            & 1180.3                            & 1171.2                            \\
                                                & 2                     & 1539.3                            & 1542.1                            & 1175.8                            \\ 
                                                & 3                     & 2893.1                            & 2666.4                            & 2660.6                            \\ 
                                                & Average ($\Bar{v}$)   & 1861.3                            & 1796.3                            & 1669.2                            \\ \hline
        \multirow{4}{*}{\ba{}}                  & 1                     & 1221.7                            & 1610.4                            & 1220.8                            \\
                                                & 2                     & 1672.9                            & 1672.6                            & 1612.4                            \\
                                                & 3                     & 2719.5                            & 2938.2                            & 2301.7                            \\ 
                                                & Average ($\Bar{v}$)   & 1871.3                            & 2073.7                            & 1711.6                            \\ \hline
        \multirow{4}{*}{\pea{}}                 & 1                     & 1300.8                            & 1278.7                            & 1296.7                            \\
                                                & 2                     & 1587.5                            & 1587.6                            & 1339.1                            \\
                                                & 3                     & 2672.8                            & 2691.4                            & 2184.8                            \\
                                                & Average ($\Bar{v}$)   & 1853.7                            & 1852.5                            & 1606.9                            \\ 
    \end{tabular}
    \end{ruledtabular}
\end{table}

\clearpage

\section{SI Note: Chiral phonons}

\subsection{Phonon circular polarization}

From the supercell approach, we obtain an atomic polarization vector $\mathbf{e}_{i, \mathbf{q}, \sigma}$ for the $i$\textsuperscript{th} atom in the unit cell at every wave vector $\mathbf{q}$ and mode index $\sigma$, which is of the form
\begin{equation}
    \label{eq:polarization_vector_1}
    \mathbf{e}_{i, \mathbf{q}, \sigma} = \left( x_{i}, y_{i}, z_{i} \right),
\end{equation}
where $x_{i}$, $y_{i}$, and $z_{i}$ represent the displacement of the $i$\textsuperscript{th} atom in the $x$-, $y$-, and $z$-direction, respectively. Thus, every eigenmode has $N$ such polarization vectors associated with it, one for every atom in the unit cell. The circular polarization of this eigenmode can determined using the following relation
\begin{equation}
    \label{eq:phonon_polarization}
    s^\alpha_{\mathbf{q}, \sigma} =  \sum_{i = 1}^{N} \mathbf{e}_{i, \mathbf{q}, \sigma}^\dagger S^\alpha \mathbf{e}_{i, \mathbf{q}, \sigma},
\end{equation}
where $S^{\alpha}$ $(\alpha = x, y, z$) are the spin-1 matrices on a Cartesian basis.

Alternatively, this can be expressed using rotation bases, which we can define for right- and left-handed motion around an axis $\alpha$. The right-handed polarization bases $\rb{\alpha}{}$ are
\begin{align}
    \label{eq:cartesian_rh_basis}
    \rb{x}{} &= \frac{1}{\sqrt{2}} ( 0, +\mathrm{i}, -1)\transpose \\
    \rb{y}{} &= \frac{1}{\sqrt{2}} (+1,  0, -\mathrm{i})\transpose \\
    \rb{z}{} &= \frac{1}{\sqrt{2}} (+1, +\mathrm{i},  0)\transpose.
\end{align}
The left-handed polarization basis $\lb{\alpha}{}$ can straightforwardly be obtained, by realizing this polarization basis is the complex conjugate of the right-handed polarization basis $\rb{\alpha}{}$ as
\begin{equation}
    \label{eq:lh_basis}
    \lb{\alpha}{} = \left( x_{j}^{*}, y_{j}^{*}, z_{j}^{*} \right)\transpose,
\end{equation}
as such resulting in the following left-handed polarization bases around the $x$-, $y$- and $z$-axis
\begin{align}
    \label{eq:cartesian_lh_basis}
    \lb{x}{} &= \frac{1}{\sqrt{2}} ( 0, -\mathrm{i}, -1)\transpose \\
    \lb{y}{} &= \frac{1}{\sqrt{2}} (+1,  0, +\mathrm{i})\transpose \\
    \lb{z}{} &= \frac{1}{\sqrt{2}} (+1, -\mathrm{i},  0)\transpose.
\end{align}
Using these rotation bases, we can express the spin-1 matrices or circular polarization operator ($\phononcircop{\alpha}$) as
\begin{equation}
    \label{eq:circular_polarization_operator}
    \phononcircop{\alpha} = \left( \rb{\alpha}{} \rbconj{\alpha}{} - \lb{\alpha}{} \lbconj{\alpha}{} \right).
\end{equation}
By putting the polarization vector in the bra-ket notation, i.e. $\ket{e_{i, \mathbf{q}, \sigma}} = \left( x_{i}, y_{i}, z_{i} \right)\transpose$, we can compute the circular polarization of a phonon eigenmode as
\begin{equation}
    \label{eq:circular_polarization_eigenmode}
    s^{\alpha}_{\mathbf{q},\sigma} = \sum^{N}_{i = 1} \bra{e_{i, \mathbf{q}, \sigma}} \phononcircop{\alpha} \ket{e_{i, \mathbf{q}, \sigma}}
\end{equation}

\subsection{Circular polarization in chiral perovskites}

\begin{figure*}[htbp!]
    \includegraphics{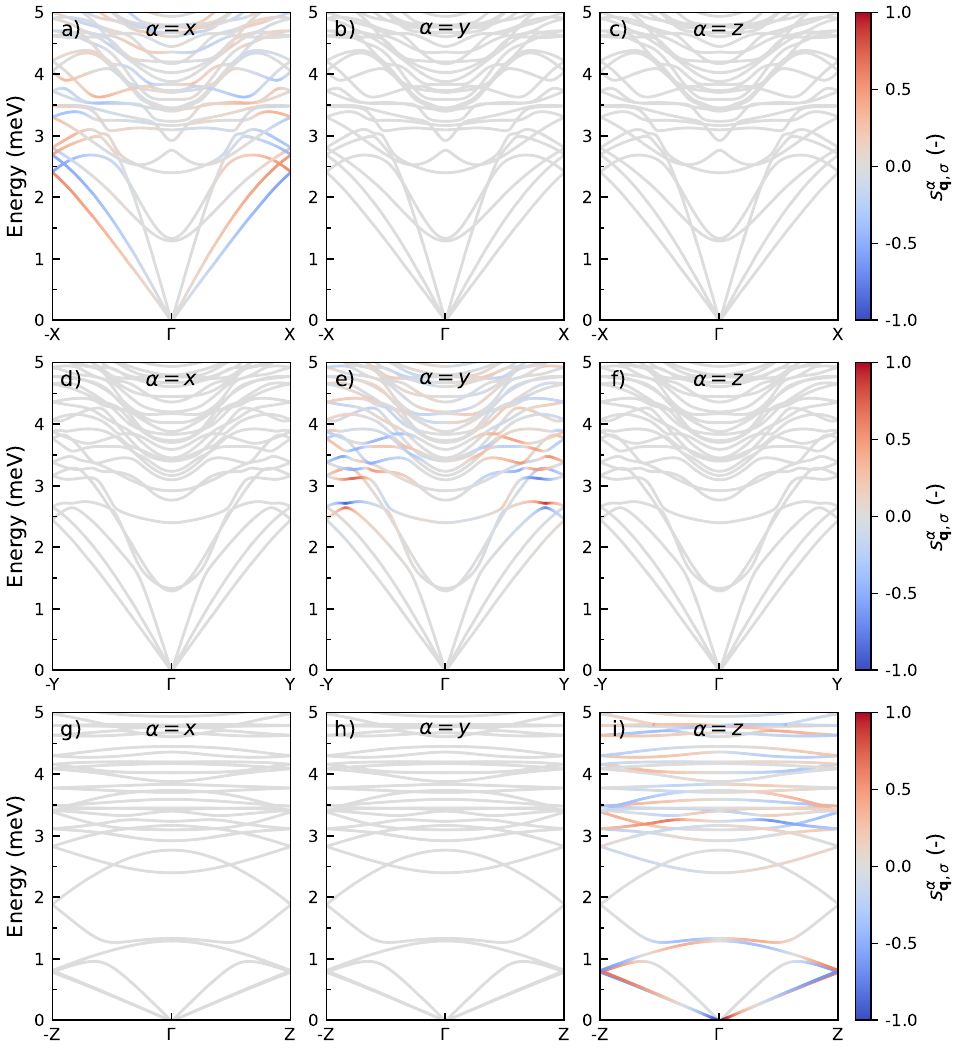}
    \caption{Circularly polarized phonon dispersion of \smbapb{}. Dispersion of phonons propagating along the (a-c) $x$-axis (Γ$-$X), (d-f) $y$-axis (Γ$-$Y), and (g-i) $z$-axis (Γ$-$Z). Phonon branches are color-coded with the circular polarization of the phonon modes, which is determined around the axis $\alpha$ = $x$, $y$, or $z$. Red, blue, and gray are used to represent right-handed ($\phononcircpolgeneral > 0$), left-handed ($\phononcircpolgeneral < 0$), and non-polarized  ($\phononcircpolgeneral = 0$) phonon modes.}
    \label{fig:chiral_phonons_orientations} 
\end{figure*}

In the main text, we only show the chirality of the phonon modes around their propagation direction (Figure 3); in other words, we only investigate if the phonon modes exhibited circular motion in the plane perpendicular to their propagation direction. Here we analyze the chirality of phonon modes of \smbapb{} in all directions. For example, for phonons propagating in the $x$-direction (Γ$-$X path) we assess their polarization around the $x$-, $y$-, and $z$-axis in Figure~\ref{fig:chiral_phonons_orientations}. For these high-symmetry paths, we observe that the phonons only possess a net circular polarization, and thus chirality, around their propagation direction, not in the directions perpendicular to their propagation direction. Thus, the phonons in the  Γ$-$X, Γ$-$Y, and Γ$-$Z paths are solely polarized around the $x$-, $y$-, and $z$-axis, respectively.

\clearpage

\subsection{Circular polarization in achiral perovskites}

\begin{figure*}[htbp!]
    \includegraphics{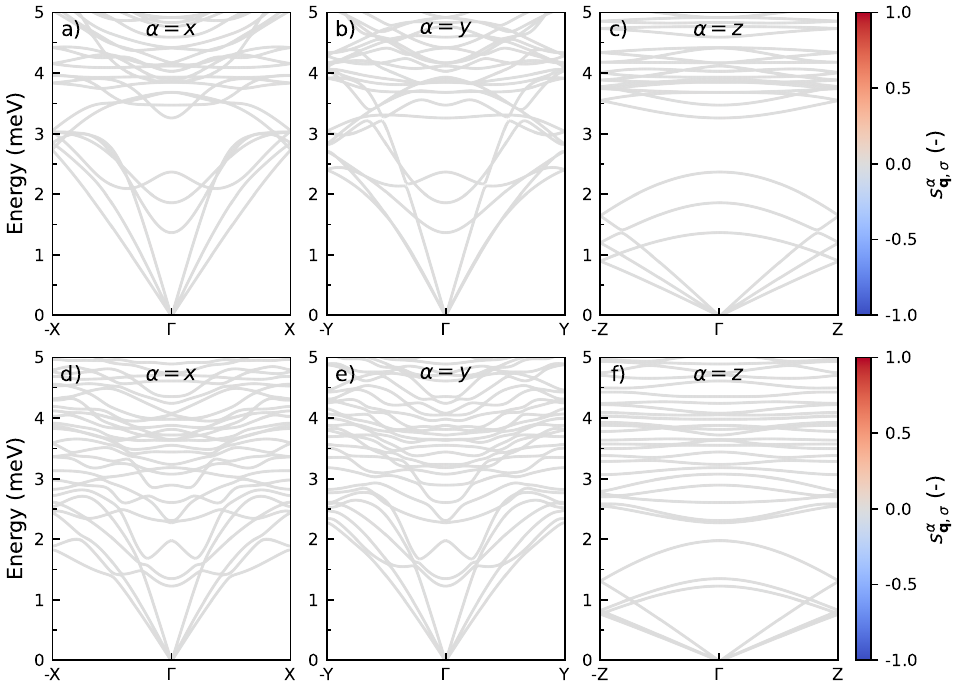}
    \caption{Circularly polarized phonon dispersion of (a-c) \ba{} and (d-f) \pea{}. Phonon branches are color-coded with the circular polarization of the phonon modes, which is determined around the axis $\alpha$ = $x$, $y$, or $z$. Red, blue, and gray are used to represent right-handed ($\phononcircpolgeneral > 0$), left-handed ($\phononcircpolgeneral < 0$), and non-polarized  ($\phononcircpolgeneral = 0$) phonon modes.}
    \label{fig:chiral_phonons_ba2pbi4_pea2pbi4} 
\end{figure*}

\begin{figure*}[htbp!]
    \includegraphics{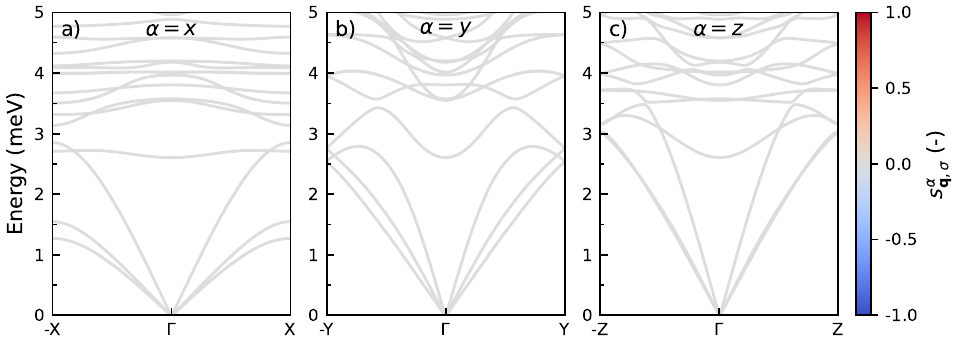}
    \caption{Circularly polarized phonon dispersion of (a-c) \racmbapb{}. Phonon branches are color-coded with the circular polarization of the phonon modes, which is determined around the axis $\alpha$ = $x$, $y$, or $z$. Red, blue, and gray are used to represent right-handed ($\phononcircpolgeneral > 0$), left-handed ($\phononcircpolgeneral < 0$), and non-polarized  ($\phononcircpolgeneral = 0$) phonon modes.}
    \label{fig:chiral_phonons_rac-mba2pbi4} 
\end{figure*}

Next, we assess the circular polarization of the phonon modes in achiral perovskites. We show the circularly polarized phonon dispersion for \ba{} and \pea{} in Figure~\ref{fig:chiral_phonons_ba2pbi4_pea2pbi4} and for \racmbapb{} in Figure~\ref{fig:chiral_phonons_rac-mba2pbi4}. To do so, we adopted the same naming convention for the special points as in the main text with the structures from Table~\ref{tab:perovskites_optimized}; X = ($\frac{1}{2}$, 0, 0), Y = (0, $\frac{1}{2}$, 0), and Z = (0, 0, $\frac{1}{2}$), with -X = (-$\frac{1}{2}$, 0, 0). The phonon dispersions show that all phonon modes are achiral in the achiral perovskites.

\clearpage

\subsection{Circular polarization spectrum}

\begin{figure*}[htbp!]
    \includegraphics{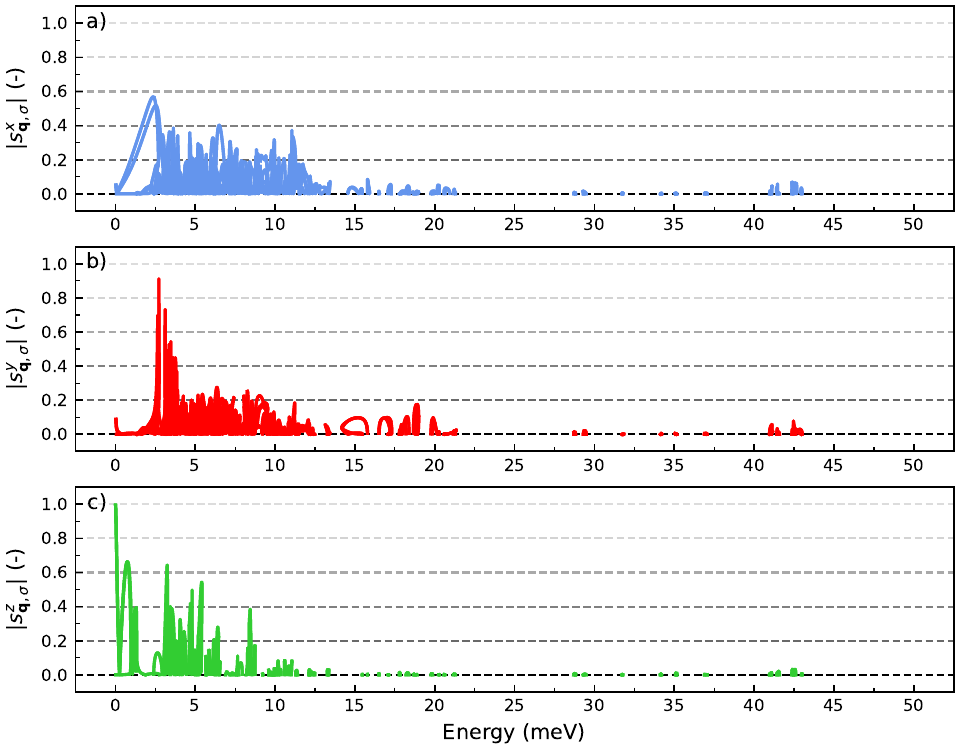}
    \caption{Circular polarization spectrum of phonons in \smbapb{}. The absolution polarization ($| \phononcircpolgeneral{} |$) spectrum of the phonons along the (a) Γ$-$X, (b) Γ$-$Y, and (c) Γ$-$Z paths. Dashed lines act as guides to the eye.}
    \label{fig:chiral_phonons_spectrum} 
\end{figure*}

To clarify what part of the phonon spectrum is chiral, we plot the chirality of the phonon modes as a function of the phonon frequency in Figure~\ref{fig:chiral_phonons_spectrum}. From this circular polarization spectrum we find that the highly chiral modes, which we characterize with ($| \phononcircpolgeneral{} | >$ 0.2) are all low energy phonons with energies below \SI{10}{\meV}.

\clearpage

\section{SI Note: Phonon modes}

\subsection{Atomic motion in eigenmodes}

As mentioned before, phonon eigenmodes are described at every wave vector $\mathbf{q}$ and mode index $\sigma$ by a polarization vector $\mathbf{e}_{i,\mathbf{q},\sigma}$ for which $i$ labels the atoms in the unit cell. This polarization vector has the form
\begin{equation}
    \label{eq:polarization_vector_2}
    \mathbf{e}_{i, \mathbf{q}, \sigma} = \left( x_{i}, y_{i}, z_{i} \right),
\end{equation}
where $x_{i}$, $y_{i}$, and $z_{i}$ denote the displacement of the $i$\textsuperscript{th} atom in the $x$-, $y$-, and $z$-direction, respectively. Noting that the polarization vectors are written in mass normalized coordinates, they can be written in terms of real-space displacements as
\begin{equation}
    \label{eq:polarization_vector_real_space}
    \Tilde{\mathbf{e}}_{i,\mathbf{q},\sigma} = \frac{1}{\sqrt{m_{i}}} \left( x_{i}, y_{i}, z_{i} \right),
\end{equation}
where $m_{i}$ is the mass of the $i$\textsuperscript{th} atom in the unit cell. The real-space displacements of the unit cell can consequently be used to determine the atomic displacements for a phonon eigenmode as
\begin{equation}
    \mathbf{u}_{i, \mathbf{q}, \sigma} \left( t \right) = \Tilde{\mathbf{e}}_{i,\mathbf{q},\sigma} \mathrm{e}^{ \mathrm{i} \left[ \mathbf{q} \cdot \mathbf{r}^{0}_{i} - \omega_{\mathbf{q},\sigma} t \right] },
\end{equation}
where $t$ is the time, $\mathbf{r}^{0}_{i}$ is the equilibrium position of the $i$\textsuperscript{th} atom, and $\omega_{\mathbf{q},\sigma}$ is the angular frequency of the phonon eigenmode. The time-dependent atomic positions follow from this as
\begin{equation}
    \label{eq:atomic_positions}
    \mathbf{r}_{i, \mathbf{q}, \sigma} \left( t \right) = \mathbf{r}^{0}_{i} + \mathbf{u}_{i, \mathbf{q}, \sigma} \left( t \right).
\end{equation}

\subsection{Eigenmodes in chiral perovskites}

\begin{figure*}[htbp!]
    \includegraphics{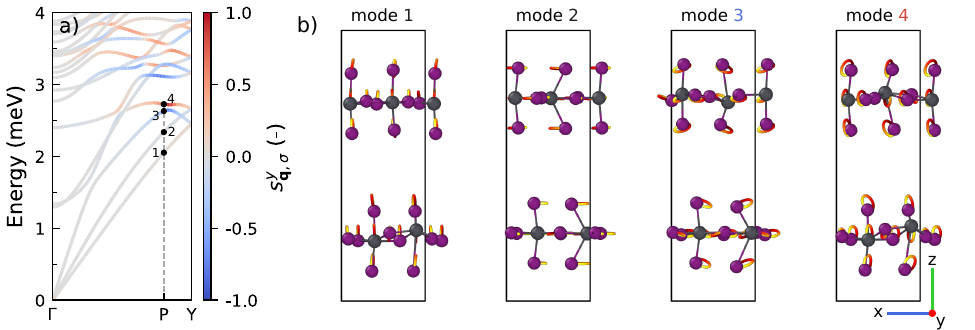}
    \caption{Eigenmodes of in-plane phonons propagating along the $y$-axis (Γ$-$Y) in \smbapb{}. (a) Phonon dispersion and (b) atomic motion in the selected phonon modes at P = (0, $\frac{2}{5}$, 0). All mode numbers are colored to indicate the circular polarization, with red (right-handed), black (non-polarized), and blue (left-handed). The atoms follow the trajectories from red to yellow as time progresses.}
    \label{fig:phonon_atomic_motion_y-axis} 
\end{figure*}

\begin{figure*}[htbp]
    \includegraphics{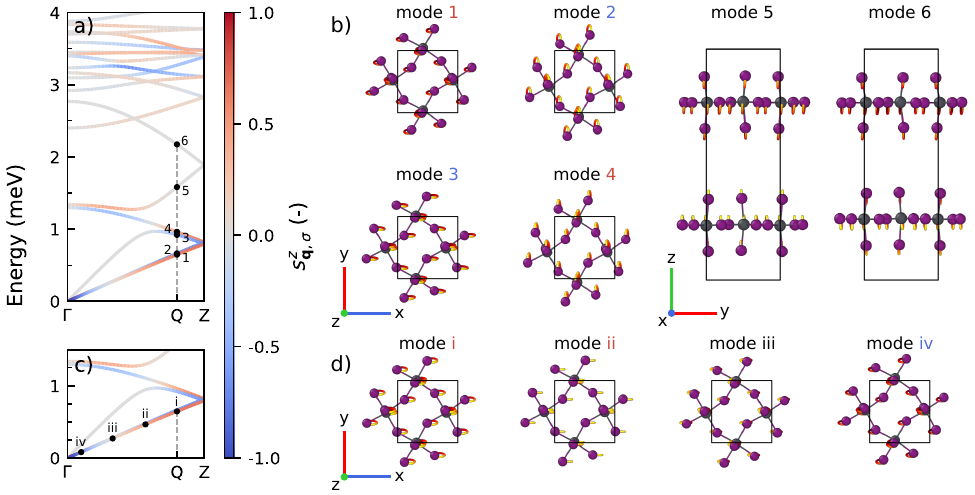}
    \caption{Eigenmodes of out-of-plane phonons (Γ$-$Z) in \smbapb{}. (a) Phonon dispersion and (b) atomic motion in the selected phonon modes at Q = (0, 0, $\frac{2}{5}$). (c) Phonon dispersion and (d) atomic motion in selected phonon modes along the Q$-$Γ path. All mode numbers are colored to indicate the circular polarization, with red (right-handed), black (non-polarized), and blue (left-handed). The atoms follow the trajectories from red to yellow as time progresses.}
    \label{fig:phonon_atomic_motion_z-axis} 
\end{figure*}

Having explored the atomic motion for phonons of \smbapb{} propagating along the $x$-axis in Figure 4, we here illustrate the atomic motion of phonons propagating along the $y$-axis (Figure~\ref{fig:phonon_atomic_motion_y-axis}) and $z$-axis (Figure~\ref{fig:phonon_atomic_motion_z-axis}). Analogous to the main text, we pick a point along the path and visualize the lowest energy modes, for the Γ$-$Y path P = (0, $\frac{2}{5}$, 0) and the Γ$-$Z path Q = (0, 0, $\frac{2}{5}$). In all chiral phonons the atoms exhibit a clear rotational motion, whereas for the achiral phonons the atoms show a linear oscillatory motion.

Analyzing the phonons propagating along the $z$-axis (Γ$-$Z), we first assess the changes in the modes upon going to phonon modes with higher energies at the Q (Figure~\ref{fig:phonon_atomic_motion_z-axis}a). The atomic motion associated with the six lowest energy modes is shown in Figure~\ref{fig:phonon_atomic_motion_z-axis}b. The lowest four phonon modes at Q are highly chiral, with all modes exhibiting an elliptical planar movement in the perpendicular plane ($xy$-plane). The two lowest modes are of opposite chirality and nearly degenerate along the full Γ$-$X path. We note that the semi-major axis of the elliptical motion of the two modes is along a different direction, i.e. $x$-axis for mode 1 and $y$-axis for mode 2. The following two modes, mode 3 and mode 4, exhibit similar characteristics. The next two modes, mode 5 and 6, exhibit a linear oscillatory motion primarily in the propagation direction and are thus achiral. 

We then investigate the effects of moving from point Q towards Γ along the lowest energy phonon branch. Along this branch we observe a change in the sign of the circular polarization of the phonon mode (Figure~\ref{fig:phonon_atomic_motion_z-axis}c), as a result of atomic motion in the phonon modes (Figure~\ref{fig:phonon_atomic_motion_z-axis}d). Close to Q the modes on this branch are highly chiral (mode i). Upon moving to Γ, we observe the phonon modes lose some of their right-handed circular polarization, as a result of more elliptical atomic motion (mode ii). The circular polarization completely vanishes when moving even further away from Q, where the atomic motion of the mode becomes linear (mode iii). Moving past the point of vanishing chirality, we observe the phonon mode attains an opposite chirality (mode iv).

\subsection{Eigenmodes in structural enantiomers}

\begin{figure*}[htbp]
   \includegraphics{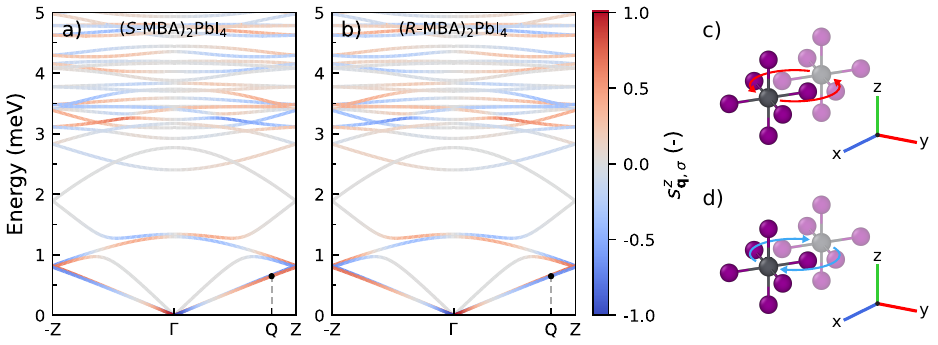}
   \caption{Circularly polarized phonon dispersion and modes of \mbapb{}. Phonon dispersion of out-of-plane (Γ$-$Z) phonons in (a) \smbapb{} and (b) \rmbapb{}. In the lowest energy mode at Q = (0, 0, $\frac{2}{5}$) the inorganic framework moves in a (c) counter-clockwise motion in \smbapb{} and (d) clockwise motion in \rmbapb{} around the $z$-axis. Red, blue, and gray are used to represent right-handed ($\phononcircpolgeneral > 0$), left-handed ($\phononcircpolgeneral < 0$), and non-polarized  ($\phononcircpolgeneral = 0$) phonon modes.}
    \label{fig:phonon_polarization_enantiomers} 
\end{figure*}

Next, to investigate the effects of structural chirality on the chirality of phonons more deeply, we compare the phonon dispersion of the two enantiomers of \mbapb{}. The phonon dispersion of \smbapb{} and \rmbapb{} are shown in Figure~\ref{fig:phonon_polarization_enantiomers}. We observe that phonons propagating in the same direction in each enantiomer have opposite polarization (Figure~\ref{fig:phonon_polarization_enantiomers}a-b); for each phonon branch, a right-handed phonon in \smbapb{} becomes a left-handed phonon in \rmbapb{} and vice versa. As an illustration, we show the motion of the inorganic layers in the lowest energy mode at Q = (0, 0, $\frac{2}{5}$) in Figure~\ref{fig:phonon_polarization_enantiomers}c-d. Whereas the inorganic layers exhibit a counter-clockwise motion around the $z$-axis in \smbapb{} (Figure~\ref{fig:phonon_polarization_enantiomers}c), a clockwise motion around the $z$-axis is observed for the equivalent mode in \rmbapb{} (Figure~\ref{fig:phonon_polarization_enantiomers}d). Interestingly, the opposite chirality of the phonon modes can be related to the structural distortions in the chiral perovskites, which between the two enantiomers has an opposite handedness, which was also observed in bulk \ch{Te}~\cite{chenChiralPhononDiode2022}. Altogether this highlights the interplay between structural chirality and the phonon handedness in such chiral materials.

\clearpage

\section{Phonon angular momentum}

\subsection{Response tensor}

In equilibrium, $\mathbf{J}^{\mathrm{ph}}$, the angular momentum per unit volume of a phonon system~\cite{zhangAngularMomentumPhonons2014} is expressed as
\begin{equation}
    \label{eq:volumetric_angular_momentum}
    \mathbf{J}^{\mathrm{ph}} = \frac{\hbar}{V} \sum_{\mathbf{q}, \sigma} \mathbf{s}_{\mathbf{q}, \sigma} \left [ f_{0} \left( \omega_{\mathbf{q}, \sigma} \right) + \frac{1}{2} \right ],
\end{equation}
where $\hbar$ is the reduced Planck constant, $V$ is the unit cell volume, $f_{0} = 1 / \left( \exp \left [ \hbar \omega_{\mathbf{q}, \sigma} / k_{\mathrm{B}} T \right ] - 1 \right)$ is the Bose-Einstein distribution, with $k_{\mathrm{B}}$ the Boltzmann constant and $T$ the temperature. The circular polarization of a phonon is provided by the vector $\mathbf{s}_{\mathbf{q}, \sigma} = (s^{x}_{\mathbf{q}, \sigma}, s^{y}_{\mathbf{q}, \sigma}, s^{z}_{\mathbf{q}, \sigma})$. As a result of time-reversal symmetry, the phonon angular momentum is an odd function in $\mathbf{q}$, i.e. $\mathbf{s}_{\mathbf{q}, \sigma} = -\mathbf{s}_{-\mathbf{q}, \sigma}$, resulting in zero sum in Equation~\ref{eq:volumetric_angular_momentum}. As proposed by \citet{hamadaPhononAngularMomentum2018}, the system can be brought out-of-equilibrium, through the application of a temperature gradient. Following Boltzmann transport theory, the distribution function deviates from the Bose-Einstein distribution as
\begin{equation}
    \label{eq:bose_einstein_linearized}
    f_{\mathbf{q}, \sigma} = f_{0} \left( \omega_{\mathbf{q}, \sigma} \right) - \sum_{\mathbf{q}, \sigma; \beta = x, y, z} \tau v^{\beta}_{\mathbf{q}, \sigma} \frac{\partial f_{0} \left( \omega_{\mathbf{q}, \sigma} \right)}{\partial T} \frac{\partial T}{\partial x^{\beta}}
\end{equation}
where $\tau$ is the phonon relaxation time, $v^{\beta}_{\mathbf{q}, \sigma}$ and $\frac{\partial T}{\partial x^{\beta}}$ are the $\beta$ component of the group velocity and temperature gradient in real space. The above approximation holds for small deviations from equilibrium, under assumption that the system relaxes back to equilibrium through phonon-phonon interactions. We employ the constant relaxation time approximation for this process, which assumes $\tau$ is independent of both $\mathbf{q}$ and $\sigma$. Combining Equation~\ref{eq:volumetric_angular_momentum} with Equation~\ref{eq:bose_einstein_linearized}, the total angular momentum per unit volume~\cite{hamadaPhononAngularMomentum2018, choiDivergencePhononAngular2022}, generated from a temperature gradient, is
\begin{equation}
    \label{eq:volumetric_angular_momentum_generation}
    J^{\mathrm{ph},\alpha} = - \frac{\hbar\tau}{V} \sum_{\mathbf{q}, \sigma; \beta=x,y,z} s^{\alpha}_{\mathbf{q}, \sigma} v^\beta_{\mathbf{q}, \sigma} \frac{\partial f_{0} \left( \omega_{\mathbf{q}, \sigma} \right)}{\partial T} \frac{\partial T}{\partial x^{\beta}} \equiv \sum_{\beta} \alpha^{\alpha \beta} \frac{\partial T}{\partial x^{\beta}},
\end{equation}
with $s^{\alpha}_{\mathbf{q}, \sigma}$ the $\alpha$ component of the phonon angular momentum and $\alpha^{\alpha \beta}$ the phonon angular momentum response tensor, with the components: $\alpha, \beta = x, y, z$.

\subsection{Symmetry constraints}

\citet{hamadaPhononAngularMomentum2018} did a symmetry analysis of $\alpha^{\alpha \beta}$, the phonon angular momentum response tensor. Through the $\mathbf{q}$ = (0, 0, 0) nature of the effect, they found that the nonzero elements of the tensor were determined by the point group symmetry of the structure, and not the space group symmetry. In case of \smbapb{}, which has the space group \triplescrew{} (point group $D2$), the response tensor takes the following form
\begin{equation}
    \label{eq:phonon_angular_momentum_response_tensor}
    \alpha^{\alpha \beta} = 
    \begin{pmatrix}
        \alpha^{xx}     & 0             & 0             \\
        0               & \alpha^{yy}   & 0             \\
        0               & 0             & \alpha^{zz}   
    \end{pmatrix},
\end{equation}
where all diagonal elements have unique values; $\alpha^{xx} \neq \alpha^{yy} \neq \alpha^{zz}$.

\subsection{Convergence}

\begin{figure*}[htbp]
    \includegraphics{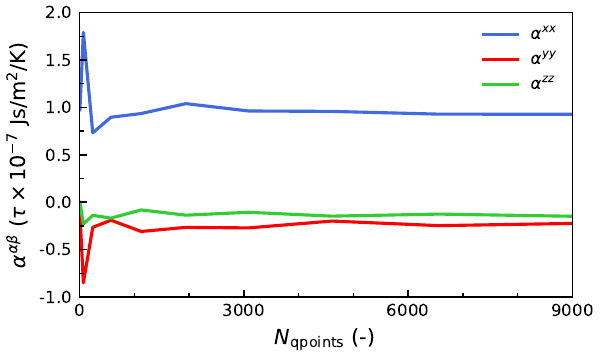}
    \caption{Convergence of the response tensors for angular momentum generation as a result of a temperature gradient at \SI{300}{\K}.}
    \label{fig:phonon_angular_momentum_convergence} 
\end{figure*}

To assess if the response tensor for angular momentum generation in the presence of a temperature gradient ($\alpha^{\alpha \beta}$) has converged, we compute the response tensor at \SI{300}{\K} for a range of reciprocal meshes. We vary these $\mathbf{q}$-meshes from \kpoints{3}{3}{1} ($N_{q}$ = 9) to \kpoints{30}{30}{1} ($N_{q}$ = 9000). In Figure~\ref{fig:phonon_angular_momentum_convergence} we show the values of $\alpha^{xx}$, $\alpha^{yy}$, and $\alpha^{zz}$ as a function of $N_{q}$, the number of points on the $\mathbf{q}$-mesh. The response tensor converges for increasingly larger reciprocal meshes, with minimal changes occurring for grids denser than \kpoints{21}{21}{7} ($N_{q}$ = 3087). Hence, the response tensors and their temperature dependence, were all computed on a \kpoints{21}{21}{7} $\mathbf{q}$-mesh in this work.

\clearpage

\bibliography{supporting}